\documentclass[10pt, conference]{IEEEtran}
\usepackage[utf8x]{inputenc} 
\usepackage{graphicx}
\usepackage{float}
\usepackage{scalefnt} 
\usepackage[table,xcdraw]{xcolor}
\usepackage{url}
\usepackage{scalefnt} 
\usepackage{verbatim}

\begin{document}

\newcommand\blfootnote[1]{%
  \begingroup
  \renewcommand\thefootnote{}\footnote{#1}%
  \addtocounter{footnote}{-1}%
  \endgroup
}

\title{Improving the network traffic classification using the Packet Vision approach}

\newif\iffinal
\finaltrue
\newcommand{\jemsid}{35}

\iffinal

\author{
\IEEEauthorblockN{Rodrigo Moreira\IEEEauthorrefmark{1}\IEEEauthorrefmark{2}, Larissa Ferreira Rodrigues\IEEEauthorrefmark{1}}
\IEEEauthorblockA{\IEEEauthorrefmark{1}Instituto de Ciências Exatas e Tecnológicas\\
Universidade Federal de Viçosa -- UFV\\
Rio Paranaíba - MG - Brasil\\
Email: \{rodrigo, larissa.f.rodrigues\}@ufv.br}
\and
\IEEEauthorblockN{Pedro Frosi Rosa\IEEEauthorrefmark{2}, Flávio de Oliveira Silva\IEEEauthorrefmark{2}}
\IEEEauthorblockA{\IEEEauthorrefmark{2}Faculdade de Computação - FACOM\\
Universidade Federal de Uberlândia -- UFU\\
Uberlândia - MG - Brasil \\
Email: \{rodrigo.moreira, pfrosi, flavio\}@ufu.br}
}

\else
  \author{WVC paper ID: \jemsid \\ \\ \\ \\}
\fi
\maketitle

\begin{abstract}
\blfootnote{\color{red} This paper has been accepted by the Workshop on Computer Vision (WVC) 2020. The definite version
of this work was published by SBC-OpenLib as part of the conference proceedings. DOI: \url{https://doi.org/10.5753/wvc.2020.13496}} 
The network traffic classification allows improving the management, and the network services offer taking into account the kind of application. The future network architectures, mainly mobile networks, foresee intelligent mechanisms in their architectural frameworks to deliver application-aware network requirements. The potential of convolutional neural networks capabilities, widely exploited in several contexts, can be used in network traffic classification. Thus, it is necessary to develop methods based on the content of packets transforming it into a suitable input for CNN technologies. Hence, we implemented and evaluated the Packet Vision, a method capable of building images from packets raw-data, considering both header and payload. Our approach excels those found in state-of-the-art by delivering security and privacy by transforming the raw-data packet into images. Therefore, we built a dataset with four traffic classes evaluating the performance of three CNNs architectures: AlexNet, ResNet-18, and SqueezeNet. Experiments showcase the Packet Vision combined with CNNs applicability and suitability as a promising approach to deliver outstanding performance in classifying network traffic.

\end{abstract}

\IEEEpeerreviewmaketitle

\section{Introduction}\label{intro}  

Classifying network traffic allows us to know the kind of application running on the network, benefiting the models for forecasting, capacity utilization, quality of service, security, and planning and management steps. Besides, in the frameworks of new communication, network architectures require intelligent entities to support resources management and operation. Traffic classification mechanisms are known and widely explored in the state-of-the-art, however, with the advent of convolutional neural networks (CNNs), new methods of training, validation, and classification are available, especially those based on images raising the opportunity to propose and evaluate mechanisms for network traffic classification \cite{8473682} \cite{lim2019payload}.

Among the known traffic classification mechanisms, we can categorize them as port-based, payload-based, machine-learning approaches based on statistics and deep learning \cite{4738466}. In particular, CNNs demonstrate capabilities beyond its fields of action with highly accurate mechanisms for clustering and classifying medical images \cite{Rodrigues2020}, biomolecular \cite{Nagao2020}, environmental \cite{Nogueira2017}, and others contexts \cite{Guo2016}. The success of CNNs is due to their ability to incorporate spatial context and weight sharing between pixels in order to extract high-level hierarchical representations of the data \cite{Ponti2017}.

In this sense, we employ the capabilities of CNNs for processing packets of data communication networks. The graphics processing supported by GPU hardware surpasses the CPU-based processing because reducing the execution time \cite{6024781}. Hence, the speedup of time-to-ready of traffic classification technologies is reducing \cite{7979887}, enabling faster classification.


Recent studies demonstrated effective results in network traffic classification using deep CNNs \cite{8473682} \cite{lim2019payload} . However, these studies performed the classification splitting both header and payload of packets as a learning feature. In a real scenario, this approach may generate security and time issues, regarding the last one, it may increase the pre-processing time without guaranteeing gains in classification performance metrics.

In this paper, we proposed the Packet Vision: a method based on computer vision to generate images from both payload and packet header. Our main contribution relies on generating a single image representing all content of the network packet. Other approaches for traffic classification, such as those based on the packet signature \cite{10.1145/1162678.1162679}, conflict with security, and privacy aspects, since information as the source and destination address, port, and transport protocol, to name a few are handling as plain text, making straightforward inference by malicious third-parties. Furthermore, a novel contribution of this paper is an evaluation of the performance of three state-of-the-art CNNs for the network traffic classification via training from scratch and fine-tuning.

Our results showcase the suitability and performance score of Packet Vision in generating and classifying images of packets from communication networks considering from raw-data. Besides, we considered three classification technologies based on CNNs, applying a hypothesis test to judge the performance between them.

The remaining of this paper is organized as follows: Section~\ref{sec:relate_work} surveys related work. Section~\ref{sec:packet_vision} presents our approach for network traffic classification. The CNNs evaluated in this paper, and the protocol used in the experiments are presented in Section~\ref{sec:classification_method}. Section~\ref{sec:resultados} presents and discusses the results. Finally, we provide concluding remarks and future work agenda in Section~\ref{sec:conclusion}.

\section{Related Work}\label{sec:relate_work}  

Lim et al. \cite{lim2019payload}, proposed a traffic classification mechanism aims to improve the quality of service for applications without interference from the network operator. Its structure generates a dataset containing images of flows analyzed over time intervals. The approach uses CNN and Long short-term memory (LSTM) to train and evaluate the classification performance using the F1-score metric. The proposed architecture considers three layers, the lowest containing data switches, including switches and hosts that exchange data between their own, on top of previous the control including classification mechanisms and traffic entities. The topmost layer allows the implementation of specific network behaviors based on the type of traffic.

The image generation mechanism for the dataset comprises capturing the flow: a set of packets with similar characteristics (source and destination host, port, and transport protocol) in a specific time interval. Therefore, for each packet of a flow, extracts its payload and performs mathematical operation over a set of bits to transform it into a single numerical value, consequently a single pixel. Thus, a single figure, containing many pixels, is the set of packet representing an application's flow. This approach does not carry out cross-validation and disregard the entire package structure, requiring additional computation in the processing step that consists of extracting the payload of each package.

Vasan et al. \cite{Vasan2020} proposed an architecture of CNN and evaluates the virtual threats as malware close classification real-time. The construction of the dataset transforms the binary signature of malware, which is an 8-bit vector into an 8-bit array, afterward in a grayscale figure and then applying a 2-D color map. The classification performance evaluation takes into account approaches with data augmentation and fine-tuning. Unlike the present proposal, this article proposes cross-validation to avoid bias and over model adjustment, besides there is no need to transform the image of the 2-D color maps dataset, maintaining performance.

Chen et al. \cite{8258054} presents an IP traffic classification framework based on CNNs named Seq2Img. This approach consists of capturing the packets of a flow and extracting its characteristics and behaviors. A probability distribution model called Reproducing Kernel Hilbert Space (RKHS) is mandatory to construct the figures for each traffic class, consisting of the network protocols and popular social networking applications. Accuracy was the performance metric held in the validation of the traffic classification model. Unlike the present paper, the authors did not validate the proposal with hold-out, and the data collection mechanism depends on a third non-open source application. On the other hand, our approach consists of an open-source collector and does not handle images as flows and does not require processing with complex mathematical models.

Wang et al. \cite{8473682} proposed a framework for classifying malicious traffic in domestic environments through home-gateway equipment containing an embedded traffic prediction mechanism. The mechanism based on CNNs is similar to ours because they take into account the figure from each package as a data suitable for Machine learning models. However, different from us in the pre-processing stage, the ethernet header of the package is removed. Besides, to avoid bias and overfitting in the training model, we applied, according to a probability distribution, we shuffle the image pixels of each class. Thus, packages containing the same source and destination address do not keep standardized pixels in predefined locations. The dataset images built from a set of packet captures of typical Internet standard applications.

Other works are known in the state-of-the-art, proposing a network traffic classification targeting security, quality of service enhancement, management, and others \cite{8713803, 8026581, 8861934}. They vary in terms of the learning and validation method, also differs between strategies based on port, payload, statistics, CNNs, flows, and others \cite{SOYSAL2010451}. The Packet Vision innovates by drawing the packets entirely, considering header and payload, and by creating a deep learning model considering those images generated through packets raw data.

\section{Packet Vision}\label{sec:packet_vision}

The resource sharing turn up in different ways in the literature. The architecture of the operating systems, especially those for time-sharing processing, has been inspiring new formats of resource sharing, impacting computing resources, and network sharing. Sharing network resource relies on to assign part of general-purpose hardware to a specific user while safeguarding essential aspects of isolation and guarantees. In the context of mobile networks, especially in the 5G standardization, sharing took the form of network slicing, which provides logical networks with independent data and control plans for users to meet specific application requirements.

Therefore, among the network slicing approaches rising the Network and Slice Orchestrator (NASOR) \cite{10.1007/978-3-030-44041-1_73} 
that implements the network slicing beyond the mobile network ecosystem, providing logical connectivity over the Internet data plane. The NASOR ecosystem includes interfaces that facilitate network slice management, called the Open Policy Interface (OPI). The OPI interface allows third-party mechanisms to support network slicing and management. Consequently, we propose a component that performs this interface, offering traffic classification to lead the NASOR path configuring agent, called Packet Vision.

\begin{figure}[!htbp]
  	\includegraphics[width=0.61\columnwidth]{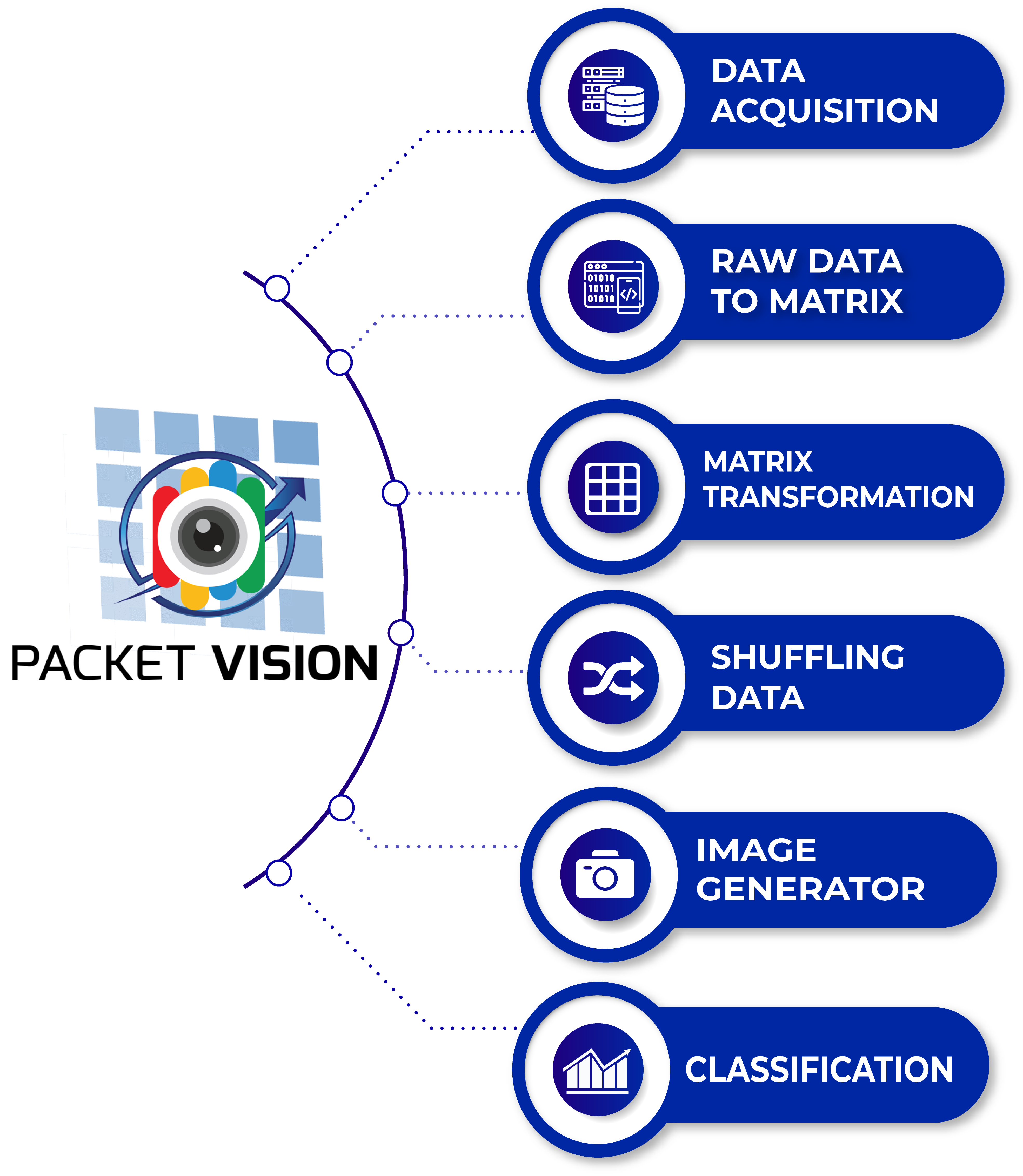}
  	\centering
	\caption{Packet Vision proposed method.}
	\label{fig:method}  
\end{figure}

The Packet Vision is a method, originated from the drawing-packet action, capable of receiving a network packet in the raw format and transforming it into images considering both the header and the payload. After generating images, it is possible to classify them according to the traffic class. Traffic classes range across the network according to the overlying application. This classification guides the network slicing agent as to the path that logical connectivity must take along Internet routers. We present Packet Vision as a method of building a dataset of network traffic class images to train and evaluate deep learning algorithms. Hence, Fig.~\ref{fig:method} depicts Packet Vision as a method for creating Datasets. 

The first step comprises collecting network packets carried over a network interface. The open-source application Wireshark and its extension libraries allow collecting packet from a network interface without affecting the application. The Packet Vision handles packets traces from four sources, collected through the open-source tool Wireshark, containing \textit{pcap} files for each traffic class. 

The first class of traffic is the standard of IoT Applications containing around 27 heterogeneous devices such as sensors and actuators \cite{7980167}. The second packet trace comprises conventional Internet applications containing DNS and BitTorrent classes \cite{10.1007/978-3-319-04918-2_10}, also available in \textit{pcap} format.

The raw information of the packet available in this dataset had been processing in order to generate figures for each class. Finally, the third packet trace refers to network slice deployed through NASOR, considering three network domains \cite{10.1007/978-3-030-44041-1_73}. 
Hence, a VoIP application providing communication between entities being in domain $A$ targeting domain $B$ communicating with voice chunks processed by codec G.711. Our method combines three packet traces making it possible to build a dataset of figures containing four traffic classes: BitTorrent, DNS, VoIP, and IoT. 

The watcher captures the packets and presents it differently; bits is the conventional form of the physical layer. However, they are grouping in formats with semantic values, such as byte array, plain text, to name a few. Hence, the second step of the method consists of handling the data in raw format, distributed in an array of bytes, and transforming them into a matrix. In this sense, our method considers the data grouping model in the Array format, which presents the packet information in hexadecimal composition.

\begin{figure}[!htb]
    \begin{center}
		\includegraphics[width=0.9\columnwidth]{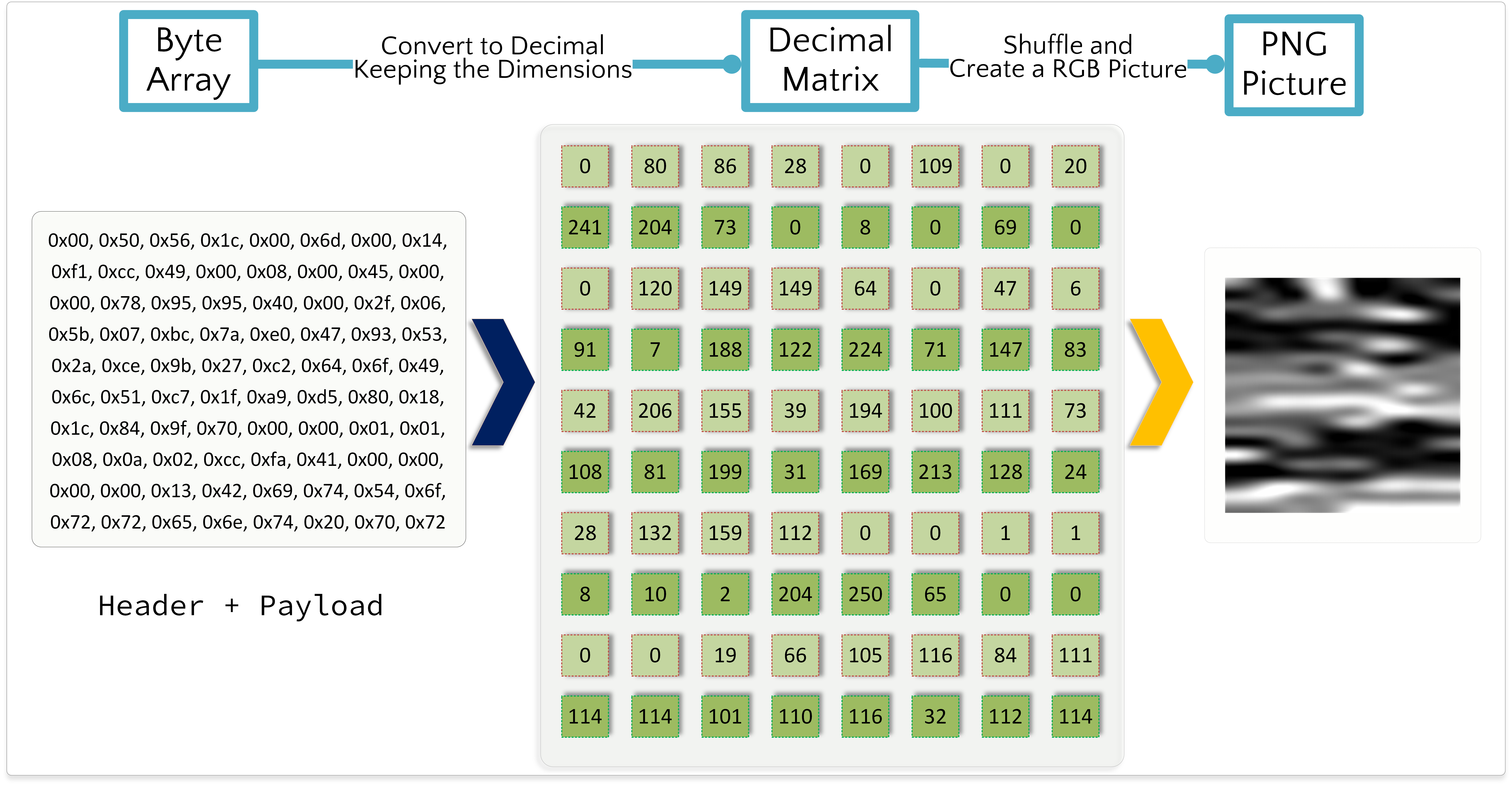}
	\end{center}
	\centering
	\caption{Building Dataset.}
	\label{fig:building_dataset}
\end{figure}

Regarding the second step, turn the hexadecimal byte array into a matrix whose size is $n \times 8$, where $n$ represents the number of rows and $8$ the number of columns according to Fig.~\ref{fig:building_dataset}. The matrix columns are 8 in size due to the native implementation of the Wireshark raw extractor library. The size of the packets, measured in bytes, varies among applications, so the method considers the number of columns fixed at $8$, and the number of rows in the matrix is variable to accommodate the size of the packet in bytes. There are scenarios where the packet size in bytes is not $n \times 8$, requiring that bytes-padding appending at the end of the packet. We agree that bytes-padding is always \textit{0xFF} for all traffic classes. Thus, when processing the matrix $n \times 8$ of hexadecimal and constructing the dataset organized in classes, these will contain figures of size $n \times 8$ pixels.

The third stage of the method considers as essential to convert the hexadecimal matrix, previously created, into decimal format. At the end of this step, the fourth shuffles the decimal values of the matrix to avoid bias and overfitting in the deep learning model. Shuffling is mandatory to change the fixed place of packet headers, such as source host, destination host, port, to name a few. Our shuffling method held Poisson probability distribution over the decimal matrix, handling the security and privacy lack in the state-of-the-art. The decimal values representing the header may remain at fixed locations in the matrix, regardless of the package content. Therefore, the fourth stage performs a shuffling of the values according to a Poison probability distribution.

The fifth step of the proposed method consists of adding RGB channels according to each decimal in the matrix, maintaining the color intensity for the three channels. The fifth step brings PNGs figures representing the contents of the packet, including the headers and the payload as an image texture. Headers are the addressing information essential to the entire packet deliver, and the payload is the information carried.

\begin{figure}[!htbp]
	\begin{center}
		\includegraphics[width=0.65\columnwidth]{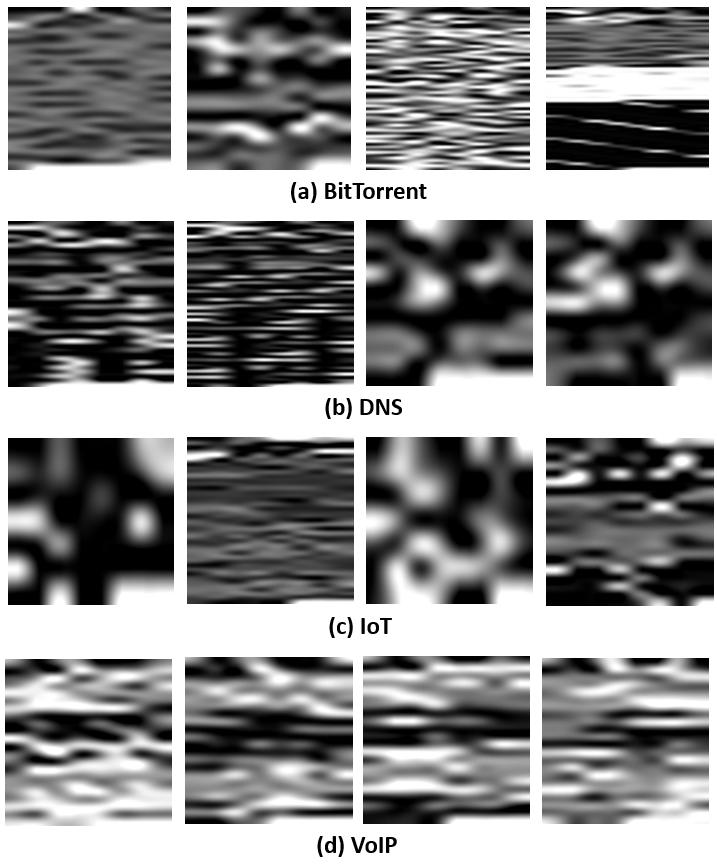}
	\end{center}
	\caption{Network packets samples generated from Packet Vision.}
	\label{fig:packet-imaged_example}
\end{figure}

The information about the created dataset has been summarizing in both Table~\ref{tab:dataset_distribution} and Fig.~\ref{fig:packet-imaged_example}, where the last one depicts examples of how Packet Vision can draw packets categorizing them into classes. This dataset are available at $<$\url{https://romoreira.github.io/packetvision/}$>$ under open-source license.

\begin{table}[!htbp]
\renewcommand{\arraystretch}{1.5}\scalefont{1.15}
\centering
\caption{Distribution of images by classes.}
\begin{tabular}{lc} \hline
\multicolumn{1}{c}{\textbf{Class}} & \textbf{Samples} \\ \hline
Bit Torrent                        & 1217             \\
DNS                                & 1412             \\
VoIP                               & 1320             \\
IoT                                & 1848             \\ \hline
Total                              & 5797        \\ \hline \hline   
\end{tabular}
\label{tab:dataset_distribution}
\end{table}

The sixth step includes training and validating the deep learning mechanism that uses the properly labeled figures from the created dataset. Many convolutional neural network architectures are known, so it is necessary to evaluate the performance of some to identify the most suitable for this kind of problem. After training and validating the learning model based on the figures generated through the raw packets, the characteristic of the current traffic on the network may be collected from a given network channel, by sample or for a determined time.

Other methods for building images from the packet are known \cite{8258054, 8845315, 8969702}, although they do not handle the complete packet structure. Alternatively, our method does not require the header and the packet payload separating in advance, causing additional processing. Besides, by shuffling the packet bytes in the matrix highlights our method regarding privacy, it is not straightforward to achieve the original semantics of the packet, including a source, destination, transport protocol port, and others from generated image.

\section{Classification Method}\label{sec:classification_method}
In this study, the classification was performed using CNNs, which uses multi-layer neural networks to learn features and classifiers in different layers, at running time, and does not require handcrafted feature extraction \cite{Goodfellow2016}. Three state-of-the-art CNN architectures were selected based on their past performance in image classification tasks: AlexNet \cite{Krizhevsky2012_AlexNet}, ResNet-18 \cite{He2015_ResNet}, and SqueezeNet \cite{Forrest2016_SqueezeNet}.

AlexNet \cite{Krizhevsky2012_AlexNet} was the champion of ImageNet Large Scale Visual Recognition Challenge (ILSVRC) 2012 and is responsible for the recent popularity of neural networks. This CNN has five convolutional layers, three max-pooling layers, two fully connected layers with a final softmax. It was a breakthrough architecture since it was the first to employ non-saturating neurons and dropout connections to prevent overfitting.

ResNet, presented in \cite{He2015_ResNet}, was the champion of ILSVRC 2015 \cite{ImageNet} and has several variations with 18 to 152 layers. This network has a series of residual blocks, each composed of several stacked convolutional layers. This configuration allows accelerating the convergence of the deep layers without overfitting. In this study, we choose to work with the ResNet-18 for the sake of simplicity.

SqueezeNet \cite{Forrest2016_SqueezeNet} has a compact architecture with approximately 50 times fewer parameters than AlexNet. This CNN reduces parameters through 1$\times$1 convolutions and eight fire modules, which performs the functions of fully connected and dense layers.

We consider two strategies of training: from scratch and fine-tuning. In training from scratch, we initialize all parameters randomly, and during the training, the values of the parameters were learned directly from the dataset in all layers \cite{Goodfellow2016}, achieving better results when compared with training based on fine-tuning since the CNN learns specific features \cite{Nogueira2017} \cite{Rodrigues2020}. The fine-tuning strategy was performed over models pre-trained on the ImageNet dataset and consists of fine-tuning the parameters in the deeper layers \cite{Ponti2017}. For both training strategies, the dimension of the last fully connected layer was four, according to the number of classes.

In order to compare the CNN architectures, we trained and tested using stratified $k$-fold cross-validation method \cite{Devijver1982}. The cross-validation was repeated five times, for each iteration, one of the training folds is chosen for the test and the others for training. Also, was taken the average accuracy, precision, recall, and f1-score, measured from the confusion matrix \cite{Duda2000}.

\section{Results and Discussion}\label{sec:resultados}

All experiments were performed on a machine with an Intel i5 3.00 GHz processor, 16 GB RAM, and a GPU NVIDIA GeForce GTX Titan Xp with 12 GB memory. The experiments were programmed using Python (version 3.6) and PyTorch \cite{PyTorch} (version 1.4) deep learning framework. 

We trained the CNN architectures using Stochastic Gradient Descent (SGD) \cite{Lecun1998} optimizer, with a learning rate of 0.001, the momentum of 0.9, batch size of 32, and 50 epochs for both, training from scratch and fine-tuning. All images were resized to 224$\times$224 pixels to adapt for the input of the CNNs evaluated. The training images had augmented through vertical and horizontal flips, rotating images around its center through randomly chosen angles of between 0$^{\circ}$ and 360$^{\circ}$. 

Our experiments aim to answer the following questions: 

\begin{enumerate}
    \item What is the highest classification performance among three evaluated CNNs?
    
    \item Considering accuracy, training from scratch, and fine-tuning, what is the most suitable training method for this dataset?
    
    \item Is the performance of pre-trained CNNs statistically equivalent?
\end{enumerate}

To assess the impact of the training from scratch and fine-tuning, we analyze the classification performance of each CNN architecture according to metrics of accuracy, precision, recall, and f1-score. Regarding the classification performance, the Tables \ref{tab:result_from-scratch} and \ref{tab:result_finetuning} presents the average 5-fold cross-validation for each CNN considering training from scratch and fine-tuning, respectively. As shown, the best performance results are achieving with the training from scratch. Consequently, the best result among the three has been obtaining by the AlexNet architecture, especially the strategy which use from scratch training.

\begin{table}[!htb]
\renewcommand{\arraystretch}{1.5}\scalefont{1.22}
\centering
\caption{5-fold average values of the performance indices for each CNN architecture training from scratch.}
\resizebox{\columnwidth}{!}{
\begin{tabular}{lcccc} \hline
\multicolumn{1}{c}{\textbf{CNN}} & \textbf{Accuracy (\%)} & \textbf{Precision (\%)} & \textbf{Recall (\%)} & \textbf{F1-Score (\%)} \\ \hline
\textit{AlexNet}                          & \textbf{100.00}              & \textbf{100.00}              & \textbf{100.00 }           & \textbf{100.00}             \\
\textit{ResNet-18}                        & 99.80             & 100.00              & 100.00            & 100.00             \\
\textit{SqueezeNet}                       & 99.60     & 99.80 & 99.60   & 99.60 \\ \hline \hline
\end{tabular}
}
\label{tab:result_from-scratch}
\end{table}

\begin{table}[!htb]
\renewcommand{\arraystretch}{1.5}\scalefont{1.22}
\centering
\caption{5-fold average values of the performance indices for each CNN architecture training with fine-tuning.}
\resizebox{\columnwidth}{!}{
\begin{tabular}{lcccc} \hline
\multicolumn{1}{c}{\textbf{CNN}} & \textbf{Accuracy (\%)} & \textbf{Precision (\%)} & \textbf{Recall (\%)} & \textbf{F1-Score (\%)} \\ \hline
\textit{AlexNet}                          & 95.40             & 96.00                & 95.80            & 95.80               \\
\textit{ResNet-18}                        & 96.40               & 96.40                & 96.80            & 96.40              \\
\textit{SqueezeNet}                       & \textbf{97.60} & \textbf{97.80}      & \textbf{97.40}  & \textbf{97.60}  \\ \hline \hline
\end{tabular}
}
\label{tab:result_finetuning}
\end{table}

Although the fine-tuning technique did not improve the performance indices compared to training from scratch, this approach requires less time to train the unfrozen layers and could be suitable in real scenarios (see Table~\ref{tab:training_time}). Thus, we compared only the pre-trained CNNs in order to identify the best model.

\begin{table}[!htb]
\renewcommand{\arraystretch}{1.5}\scalefont{1.1}
\centering
\caption{Average training time for each CNN architecture considering both training strategies.}
\begin{tabular}{lcc} \hline
                    & \multicolumn{2}{l}{\textbf{Training Time (minutes)}}                                 \\ \hline
\textbf{CNN}        & \multicolumn{1}{l}{\textbf{From-scratch}} & \multicolumn{1}{l}{\textbf{Fine-tuning}} \\
\textit{AlexNet}    & 16.21                                     & 06.21                                    \\
\textit{ResNet-18}  & 37.00                                     & 14.13                                    \\
\textit{SqueezeNet} & 27.41                                     & 12.29 \\ \hline \hline                                  
\end{tabular}
\label{tab:training_time}
\end{table}

According to the results presented in Table~\ref{tab:training_time}, although training from scratch achieves high accuracy, the most suitable for this dataset considering the impact of computational cost is the SqueezeNet architecture trained with fine-tuning. Since, in real network traffic classification scenarios, approaches with lower computational cost are more appropriate.

To assess the performance, we carried \textit{Z}-Test with $95\%$ of confidence over samples of Table~\ref{tab:performance_cnns_z-test}, which contains accuracy obtained from each test set. Thus, considering AlexNet and ResNet-18, we raise the following hypotheses: $H_{0}$ -- the performance of AlexNet is equal to or less than ResNet-18. On the other hand, $H_{a}$ -- the performance of AlexNet is higher than ResNet-18. Considering the sample space of size five, we can infer the observed $Z_{obs.}$ is lower than $Z_{crit.}$, leading us to accept $H_{0}$, implying that the performance of AlexNet is equal to or less than ResNet-18.

\begin{table}[!htb]
\renewcommand{\arraystretch}{1.5}\scalefont{1.1}
\centering
\caption{5-fold test accuracy for each CNN architecture training with fine-tuning.}
\begin{tabular}{cccc} \hline
\multicolumn{1}{c}{\textbf{\textit{Fold}}} & \textbf{\textit{AlexNet (\%)}} & \textbf{\textit{ResNet-18 (\%)}} & \textbf{\textit{SqueezeNet (\%)}}  \\ \hline
1 & 93.00 & 95.00  & 96.00   \\
2 & 97.00 & 97.00  & 98.00   \\
3 & 98.00 & 98.00  & 98.00  \\   
4 & 93.00 & 95.00  & 97.00   \\ 
5 & 96.00 & 97.00 & 99.00   \\\hline \hline
\end{tabular}
\label{tab:performance_cnns_z-test}
\end{table}

Besides, we infer the performance of ResNet-18 and SqueezeNet, raising two hypotheses, namely $H_{0}$ -- the performance of ResNet-18 is less than or equal to SqueezeNet. At the same time, $H_{a}$ -- the performance of ResNet-18 is higher than SqueezeNet. Considering a sample space with size five, and a normal distribution, the observed $Z_{obs.}$ is outside the critical region, which leads us to accept $H_{0}$, implying that ResNet-18 is less than or equal to SqueezeNet.

Hence, SqueezeNet architecture pre-trained with ImageNet had been performed better than or equal to its peers. These results suggest the suitability of Packet Vision to act as a traffic classifier mechanism and, eventually, enabling its embodiment on low-cost hardware such as Raspberry Pi.

\begin{figure}[!htb]
	\begin{center}
		\includegraphics[width=1\columnwidth]{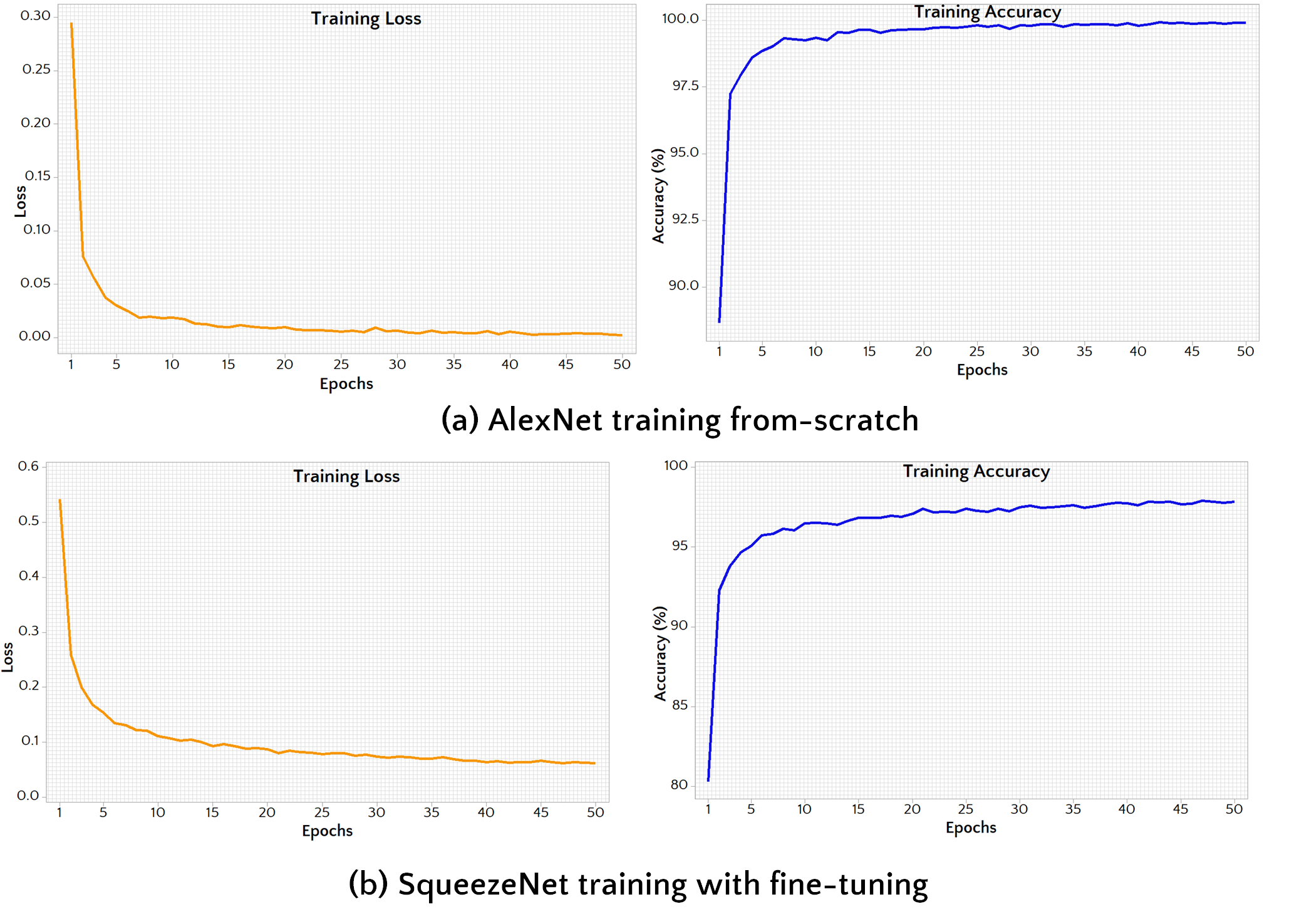}
	\end{center}
	\caption{Average 5-fold training loss and accuracy considering the best training strategy. (a) AlexNet training from-scratch; and (b) SqueezeNet training with fine-tuning.}
	\label{fig:loss_and_accuracy_cnns}
\end{figure}

Finally, considering the best result for each training strategy (from-scratch and fine-tuning), the charts in Fig.~\ref{fig:loss_and_accuracy_cnns} show how each CNN architecture behaved during the training stage, considering the average loss and accuracy of the $5$-folds. The results show that CNNs maintained the generalization property.

\section{Concluding Remarks}\label{sec:conclusion}

This paper presents the Packet Vision method for building and evaluating datasets representing traffic on communication networks through CNNs. This method allows representing the raw-data of network packets in images for training and classification in a deep learning mechanism. The image creation mechanism considering the header and the payload advances the state-of-the-art since its peers consider only the payload, among other approaches such as the semantic and statistical representation of flows. Besides, our approach is suitable for classifying traffic with similar characteristics implying in challenging tasks, achieving excellent performances according to state-of-the-art metrics, and its implementation in the network being direct by handling the packets as they are.

Carried experiments showcase that SqueezeNet performance is at least equal or higher against AlexNet and ResNet-18 trained with fine-tuning, enabling us to answer questions about the quality of CNNs performance. Besides, we point out training approaches suitability for this problem, including a statistical test seeking possible performance equivalence. Also, unlike the approaches found in the state-of-the-art, the Packet Vision shuffling step enhances the privacy claim upon packets, avoiding fixed fields of the packets at the same pixel location, avoids rebuilding the original packet from the image.

We believe that Packet Vision is a robust application for the traffic network classification with a significant degree of innovation stemming from computer vision techniques applying to generate images from packets raw-data. Moreover, the Packet Vision seems suitable for future networks, such as 5G and beyond, whose take into account the security, privacy, and application-aware as a baseline.

As future work, we intend to exploit the Packet Vision approach to generate other traffic classes related to distinct applications, such as Remote Desktop Protocol (RDP), SSH, and social media. We are also planning to evaluate other CNN architectures, data augmentation strategies, and hyperparameter optimization.

\section*{Acknowledgments}
We gratefully acknowledge the support of NVIDIA Corporation with the donation of the TITAN Xp GPU used for this research. And also, this study was financed in part by the Coordenação de Aperfeiçoamento de Pessoal de Nível Superior - Brasil (CAPES) - Finance Code 001.

\bibliography{referencias}

\end{document}